\documentclass[12pt]{iopart}

\begin{document}

\title[Critical behaviour of driven bilayer systems]
{Critical behaviour of driven bilayer systems: \\
A field-theoretic renormalisation group study}
\author{U C T\"{a}uber$^1$, B Schmittmann$^1$, and R K P Zia$^{1,2}$}

\date{\today }

\begin{abstract}
We investigate the static and dynamic critical behaviour of a uniformly
driven bilayer Ising lattice gas at half filling. Depending on the strength
of the interlayer coupling $J$, phase separation occurs across or within the
two layers. The former transitions are controlled by the universality class
of model A (corresponding to an Ising model with Glauber dynamics), with
upper critical dimension $d_{c}=4$. The latter transitions are dominated by
the universality class of the standard (single-layer) driven Ising lattice
gas, with $d_{c}=5$ and a non-classical anisotropy exponent. These two
distinct critical lines meet at a non-equilibrium bicritical point which
also falls into the driven Ising class. At all transitions, novel couplings
and dangerous irrelevant operators determine corrections to scaling.
\end{abstract}

\address{$^1$Center for Stochastic Processes in Science and Engineering, Physics Department \\ 
	Virginia Polytechnic Institute and State University, \\ 
	Blacksburg, VA 24061-0435, USA\\
	$^2$Fachbereich Physik, Universit\"{a}t - Gesamthochschule Essen,\\
	D-45117 Essen, Federal Republic of Germany.}

\jl{1}


%
%

\emph{Introduction.} Driven diffusive lattice gases (DDLG), introduced by
Katz et al. \cite{kls} to investigate \emph{far-from-equilibrium} properties
of interacting many-particle systems, are deceptively simple generalisations
of the familiar equilibrium Ising lattice gas \cite{ising}. Particles
diffuse on a lattice, controlled by not only the usual interparticle
attraction and thermal bath (at temperature $T$), but also a \emph{uniform
external force}. In conjunction with periodic boundary conditions, the
latter drives the system into a non-equilibrium steady state with
non-trivial current. A remarkable range of novel collective phenomena
emerges, some aspects of which are now well understood while many others
remain mysterious \cite{rev}. In particular, when further `slight'
modifications or generalisations are introduced, simulations often
contradict equilibrium-based expectations. In this sense, an intuitive
understanding of non-equilibrium steady states is still lacking.

While most simulations of the original model were carried out on a single
two-dimensional ($d=2$) square lattice, Monte Carlo studies have been
reported recently for two coupled driven Ising lattices stacked to form a
bilayer structure \cite{am,Hill,jsw}. Upon tuning the (Ising) \emph{%
inter-layer} coupling, the system makes a transition from a homogeneous high
temperature phase into \emph{two} distinct ordered states at low
temperatures. A simple phase diagram was found \cite{Hill}, displaying two
lines of continuous transitions which meet a line of `first-order'
transitions at a `bicritical' point. The critical properties associated with
the two lines were conjectured. In the most recent study \cite{jsw},
anisotropic \emph{intra-layer} couplings were introduced, and certain
critical properties were measured. In this letter, we report results from a
field-theoretic renormalisation group (RG) study for the continuous
transitions of this model. Even though we essentially confirm the original
conjecture \cite{Hill}, some novel and curious features emerge near the
bicritical point and along the line associated with repulsive inter-layer
couplings. After a brief description of both the microscopic model and the
field-theoretic description, we present our results.

In all simulation studies, the `microscopic' model consists of two fully
periodic $L_1\times L_2$ square lattices, arranged in a bilayer structure
(effectively, an $L_1\times L_2\times 2$ system). The sites, labeled by $%
(j_1,j_2,j_3)$, with $j_{1,2}=1,..,L_{1,2}$ and $j_3=1,2$, may be empty or
occupied. Thus, the set of occupation numbers $\{n(j_1,j_2,j_3)\}$, where $%
n=0$ or $1$, completely specifies a configuration. To access critical
points, we use half-filled systems, i.e., $\sum n=L_1L_2$. The particles
interact, so that the Hamiltonian is given by $\mathcal{H}\equiv -J_0\sum
nn^{\prime }-J\sum nn^{\prime \prime }$, where $n$ and $n^{\prime }$ are
nearest neighbours \emph{within} a given layer, while $n$ and $n^{\prime
\prime }$ differ only in the layer index. All studies focus on attractive
intra-layer interactions, $J_0>0$, while $J$ can be of either sign. Since
intra-layer anisotropies generate no qualitatively new features, $J_0$ may
be set to unity. The equilibrium phase diagram in the $J$-$T$ plane is
easily obtained, with some exactly known features. For example, a
second-order transition occurs at $T_O\simeq 0.5673/k_B$ and $J=0$ \cite
{Onsager}.

To access the most interesting \emph{non-equilibrium} properties, a \emph{%
conserved dynamics} must be imposed. Typically, Kawasaki spin exchange with
Metropolis rates is employed, i.e., particles hop to nearest neighbour holes
with probability $\min \{1,\exp {(-\Delta \mathcal{H}/k_BT)}\}$, where $%
\Delta \mathcal{H}$ is the energy change associated with the move. To model
the effects of the drive, we add $\pm E_o$ to $\Delta \mathcal{H}$ for hops
against/along, say, the 1-axis \cite{kls}, interpreting the particles as
`charged' in the presence of an external `electric' field $(E_o,0,0)$. Note
that, with Metropolis rates, it is possible to study the `infinite' $E_o$
case: jumps against the field are simply never executed. When so driven, the
phase diagram in the $T$-$J$ plane can be found (schematically shown in
Fig.~1 \cite{am,Hill,jsw}). At high $T$, the system is disordered (D). At
low $T$, the system phase separates:\ For sufficiently repulsive $J$, the
fully ordered ($T=0$) state displays densities $1$ and $0$ in the two
layers, so that this phase is labeled `full-empty' (FE). On the other hand,
for $J>0$ and low $T$, each layer phase separates individually, resulting in
strips of particles (of width $L_2/2$, at $T=0$) `on top of each other', and
aligned with the drive. Thus, we label this state the `strip' (S) phase. In
the following, we invoke field-theoretic renormalisation group techniques,
to investigate \emph{universal} properties associated with the critical
points.


\begin{figure}[tbp]
\input{epsf}
\par
\begin{center}
\begin{minipage}{.5\textwidth}
  \epsfxsize = \textwidth \epsfysize = .9\textwidth \hfill
  \epsfbox{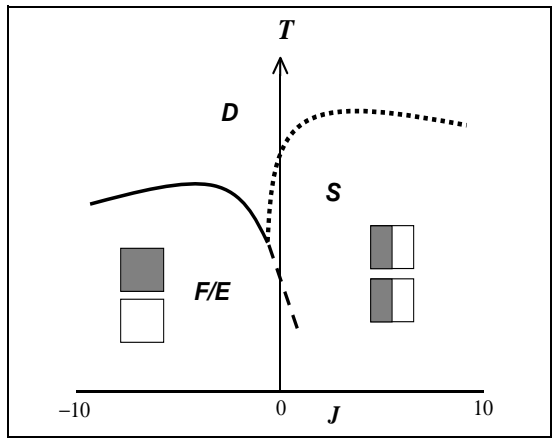}
    \vspace{-1.cm}
\end{minipage}
\end{center}
\caption{Schematic phase diagram in the $J,T$ plane, for $E_{o}=\infty $.
The D-FE (D-S) transition line is shown solid (dotted). The S and FE phases
are separated by a (dashed) first order line. The junction of the three
lines marks the bicritical point.}
\label{p1b}
\end{figure}

\emph{Model equations.} This analysis has already been initiated in \cite
{rev,shz}. We define the single-layer magnetisations $\varphi _{i}(\vec{x})$%
, $i=1,2$ as the coarse-grained versions of $2n(j_{1},j_{2},i)-1$, with the
in-layer coordinate $\vec{x}$ generalized to $d$ dimensions. To insure the
proper $E_{o}=0$ limit, we construct a Landau-Ginzburg-Wilson (LGW)
Hamiltonian, $\mathcal{H}_{c}$, which contains all terms, up to fourth order
in $\varphi _{i}$ and second order in $\vec{\nabla}\varphi _{i}$, compatible
with stability requirements and symmetries of the microscopic model. As
usual, the explicit relationships \cite{shz} between the coarse-grained
couplings and the microscopic parameters $\left( J,J_{0},T\right) $ are not
needed. The next step is to incorporate the dynamics and the
(coarse-grained) drive $E$ ($\propto \tanh E_{o}$), following Ref.~\cite{rev}%
. The result is a set of Langevin equations for the fields $\varphi _{i}(%
\vec{x},t)$. Focusing on the deterministic evolution first, the basic
equation for $\varphi _{1}$ (with a similar one for $\varphi _{2}$) reads $%
\partial _{t}\varphi _{1}=\frac{\lambda }{2}\left( 1-\varphi _{1}\,\varphi
_{2}\right) \left[ \delta \mathcal{H}_{c}/\delta \varphi _{2}-\delta 
\mathcal{H}_{c}/\delta \varphi _{1}\right] +\gamma \left[ \nabla
^{2}\,\left( \delta \mathcal{H}_{c}/\delta \varphi _{1}\right) +\sqrt{2}%
\,E\,\partial \varphi _{1}^{2}\right] $. Here, the first term models the
energetics of \emph{inter-plane} jumps, with a relaxation constant denoted
by $\lambda $. The second term reflects the (conserved) \emph{in-plane}
dynamics, being the coarse-grained version of the usual (single-layer) DDLG,
with diffusion coefficient $\gamma $. The symbol $\partial $ indicates a
spatial gradient along the direction of the drive. Some numerical factors
appear for later convenience. Anticipating the very different roles played
by the total and `staggered' magnetisations, we introduce $\Sigma (\vec{x}%
,t)\equiv [\varphi _{1}(\vec{x},t)+\varphi _{2}(\vec{x},t)]/\sqrt{2}$ and $%
\Delta \equiv [\varphi _{1}-\varphi _{2}]/\sqrt{2}$. Finally, we incoporate
the appropriate Langevin noise terms into the equations of motion, so that
the starting point of our field theoretic analysis is:

\begin{eqnarray}
&&\partial _t\Sigma =\gamma \,\nabla ^2\left[ \left( r_\Sigma -\nabla
^2\right) \Sigma +\frac u6\,\Sigma ^3+\frac{\widetilde{u}}2\,\Sigma \,\Delta
^2\right]  \label{Sig} \\
&&+\gamma \,\partial \left( \frac E2\,\Sigma ^2+\frac{\bar{E}}2\,\Delta
^2\right) +\zeta _\Sigma \ ,  \nonumber \\
&&\partial _t\Delta =-\lambda \left[ \left( r_\Delta -\nabla ^2\right)
\Delta +\frac g6\,\Delta ^3+\frac{\widetilde{g}}2\,\Delta \,\Sigma ^2\right]
\label{Del} \\
&&+\lambda \left( \widetilde{E}\,\Delta \,\partial \Sigma +\widetilde{E}%
^{\prime }\,\Sigma \,\partial \Delta \right) +\zeta _\Delta \ ,  \nonumber
\end{eqnarray}
with stochastic white noise characterised by $\langle \zeta _\Sigma \rangle
=0=\langle \zeta _\Delta \rangle $, and correlations 
\begin{eqnarray}
&&\langle \zeta _\Sigma (\vec{x},t)\,\zeta _\Sigma (\vec{x}^{\prime
},t^{\prime })\rangle =-2\gamma \,\nabla ^2\,\delta (\vec{x}-\vec{x}^{\prime
})\,\delta (t-t^{\prime })\ ,  \label{Sns} \\
&&\langle \zeta _\Delta (\vec{x},t)\,\zeta _\Delta (\vec{x}^{\prime
},t^{\prime })\rangle =2\lambda \,\delta (\vec{x}-\vec{x}^{\prime })\,\delta
(t-t^{\prime })\ .  \label{Dns}
\end{eqnarray}
Some comments are in order. First, $\bar{E}=\widetilde{E}=\widetilde{E}%
^{\prime }=E$ in mean-field, but suffer different renormalisations, so that
our parameter space must be generalized. Second, for zero drive, the D-FE
(D-S) transitions are marked by the vanishing of $r_\Delta $ $(r_\Sigma )$,
i.e., $\Delta $ $\left( \Sigma \right) $ becoming critical. Third, the local 
$\Sigma $ is conserved and unaffected by inter-layer jumps; hence, its
relaxation is diffusive and independent of $\lambda $. Indeed, for $%
\widetilde{u}=0$ and vanishing drive, Eq.~(\ref{Sig}) reduces to that in
model B dynamics \cite{HH}. Now, for non-zero drive, the terms containing
only $\Sigma $-fields are precisely the standard Langevin equation for the
DDLG \cite{JS,LC}. Therefore, Eq.~(\ref{Sig}) is supplemented with a \emph{%
conserved} stochastic noise term. On the other hand, for $\widetilde{g}=0$
and zero drive~(\ref{Del}) reduces to the purely dissipative dynamics of
model A \cite{HH}, controlled only by the inter-plane jump rate $\lambda $.
Thus, \emph{non-conserved} white noise has been added. The \emph{cross }%
correlations $\langle \zeta _\Sigma \,\zeta _\Delta \rangle $ vanish, as a
consequence of the statistical independence of intra- and inter-layer jumps.
Finally, the drive will induce the standard DDLG spatial anisotropies, to be
discussed next.

\emph{The D-S critical line. }We begin with the D-S transitions, where the
conserved field $\Sigma $ becomes massless, while $\Delta $ remains
non-critical. In the presence of the driving term, the characteristic
anisotropies of DDLG's emerge. The system softens only in the spatial sector 
\emph{transverse} to the drive, so that transverse and longitudinal momenta
scale differently: $q_{\parallel }\sim $ $q_{\perp }^{1+\kappa }$, where $%
\kappa $ is known as the anisotropy exponent ($\kappa =1$ in mean-field) 
\cite{JS,LC}. Consequently many longitudinal non-linear terms become
irrelevant. Next, since the non-conserved field $\Delta $ is massive, loop
diagrams containing internal $\Delta $ propagators are \emph{infrared}%
-convergent and may be ignored. Thus $g$ and, remarkably, $\zeta _{\Delta }$
may be neglected. Alternatively, the same conclusions can be reached by
focusing on \emph{ultraviolet} singularities in perturbation theory (see
Ref.~\cite{mt}). The result is that, within the $\Sigma $ sector, the
analysis is identical to the standard single-layer DDLG case \cite{JS,LC}.
These considerations can be summarized in the effective theory, 
\begin{eqnarray}
&&\partial _{t}\Sigma =\gamma \left[ c\,\partial ^{2}+\nabla _{\perp
}^{2}\left( r_{\perp }-\nabla _{\perp }^{2}\right) \right] \Sigma +\gamma \,%
\frac{E}{2}\,\partial \Sigma ^{2}+\zeta _{\Sigma }\ ,  \label{cds} \\
&&\partial _{t}\Delta =-\lambda \left( r_{\Delta }-\nabla ^{2}\right) \Delta
-\lambda \,\frac{\widetilde{g}}{2}\,\Delta \,\Sigma ^{2}+\lambda \,%
\widetilde{E}\,\Delta \,\partial \Sigma +\zeta _{\Delta }\ ,  \label{cdd}
\end{eqnarray}
where $c>0$ is introduced to account for anomalous anisotropy \cite{JS,LC}.
The D-S line itself is associated with a vanishing (renormalized) $r_{\perp }
$, while $r_{\Delta }>0$ (so $\nabla ^{2}\Delta $ is also irrelevant).
Denoting a momentum scale by $\mu $ ($\sim q_{\perp }$), a consistent set of
naive dimensions emerges: $\gamma \sim \mu ^{0}$, $\omega \sim \mu ^{4}$, $%
\lambda \sim \mu ^{2}$, and $c\sim \mu ^{0}$. As in the standard case, the
most relevant coupling is $\mathcal{E}=E^{2}/c^{3/2}$ ($\sim \mu ^{5-d}$
naively), so that the upper critical dimension is $d_{c}^{\Sigma }=5$. In
addition to $c$, only the coupling $\sigma =\widetilde{g}c/E\widetilde{E}$
exhibits a non-trivial flow under the RG. The remaining effective
non-linearities all have lower scaling dimension: $u,\widetilde{u},%
\widetilde{E}\bar{E},\widetilde{E}^{\prime }\bar{E}\sim \mu ^{3-d}$, and $%
\widetilde{E}\widetilde{E}^{\prime }\sim \mu ^{1-d}$. Of course, the static
coupling $u$ drives the phase transition and therefore constitutes a
dangerously irrelevant variable\cite{JS}.

Remarkably, the analysis can be performed to all orders in $\varepsilon
\equiv d_{c}^{\Sigma }-d$. In particular, the $\Sigma $-loops which modify
the new vertex ($\widetilde{E}$) are the \emph{same} as those affecting $c$.
Thus, the same power series, $\rho (\mathcal{E})$, enters the RG flow
equations for both renormalised couplings: $\beta _{\mathcal{E}}\equiv \mu
\partial _{\mu }\mathcal{E}=-\mathcal{E}[\varepsilon +\frac{3}{2}\rho (%
\mathcal{E})]$ and $\beta _{\sigma }\equiv \,\mu \partial _{\mu }\sigma
=\sigma (1+\sigma )\rho (\mathcal{E})$. For $d<5$, the stable RG fixed point
turns out to be $\rho (\mathcal{E}^{*})=-2\varepsilon /3$ and $\sigma ^{*}=-1
$ to all orders, whereas for $d\geq 5$, both couplings tend to the Gaussian
fixed point $0$. Notice, however, that the novel bilayer coupling $\sigma $
does not enter any singular diagram for the two-point functions.
Consequently, despite its non-trivial fixed-point value, it does not affect
the scaling exponents along the critical $\Sigma $ line, which are just
those of the standard DDLG. Even for $d<5$, the only non-trivial exponent is 
$\kappa $. The two-point correlation function for the $\Sigma $ fields
scales as $C_{\Sigma \Sigma }(q,\omega ,r_{\perp })\,=\,q_{\perp }^{-6}\ {%
\hat{C}}_{\Sigma }\left( q_{\perp }r_{\perp }^{-1/2},\,q_{\parallel
}/q_{\perp }^{1+\kappa },\,\omega /q_{\perp }^{4}\right) $, corresponding to
the transverse critical exponents $\eta _{\perp }=0$, $\nu _{\perp }=1/2$,
and $z_{\perp }=4$. As a consequence of Galilean invariance \cite{JS}, the
exponent $\kappa $ is fixed by a scaling relation to $1+\varepsilon /3$ for $%
2\leq d\leq 5$ ($\kappa =1$ for $d\geq 5$).

\emph{The bicritical point.\ }Remarkably, these features remain valid at the
point where both fields are critical: $r_{\Delta }=r_{\perp }=0$. Formally,
this follows from the observation that the above scaling dimensions still
hold, and no additional diagrams appear. The conserved field $\Sigma $
essentially slaves the non-conserved field $\Delta $. The larger critical
dimension $d_{c}^{\Sigma }=5$ still dominates the scaling behaviour, keeping
the static couplings $u,\widetilde{u},g$, and $\widetilde{g}$ irrelevant
(with $\tilde{g}$ appearing merely through $\sigma $). Thus, there are no
non-trivial renormalisations in the two-point function for the $\Delta $%
-fields, so that its scaling is just \emph{isotropic} and \emph{mean-field}%
-like, $C_{\Delta \Delta }(q,\omega ,r_{\Delta })\,=\,q^{-4}\ {\hat{C}}%
_{\Delta }\left( q\,r_{\Delta }{}^{-1/2},\,\omega /q^{2}\right) $.\ This
corresponds to a non-conserved Gaussian theory: $\eta =0$, $\nu =1/2$, and $%
z=2$ near five dimensions. 

\emph{The D-FE critical line.} Finally, we turn to the case where only the $%
\Delta $ field is critical: $r_{\Delta }\to 0,$ $r_{\perp }>0$. Though $%
\Sigma $ remains massive, it is a conserved field and hence a slow variable.
This situation is reminiscent of models E or G in equilibrium critical
dynamics \cite{HH}. The key difference resides, however, in the external
drive rendering the system far-from-equilibrium and manifestly anisotropic.
At the same time, the coupling to the diffusive mode sets us outside the
framework of the simple nonequilibrium kinetic Ising models investigated in~%
\cite{gjh}. So, a full RG calculation is required to determine whether
model-A relaxational kinetics still persists.

With $\Sigma $ non-critical, we may safely neglect the terms $\nabla
^{4}\Sigma $ and $\nabla ^{2}\Sigma ^{3}$. Naively, scaling is isotropic
along this critical line, so that $(q_{\parallel },q_{\perp })\sim \mu $, $%
\omega \sim \mu ^{2}$, and $(\lambda ,\gamma )\sim \mu ^{0}$. The dominant
non-linear coupling is now $g\sim \mu ^{4-d}$, with an upper critical
dimension $d_{c}^{\Delta }=4$. Power counting yields $\widetilde{u}\sim \mu
^{2-d}$ which therefore becomes irrelevant. However, beyond $g$, several
additional marginal effective couplings appear, consisting of combinations
such as $E\bar{E}$, $\widetilde{E}\bar{E}$, $\widetilde{E}^{\prime }\bar{E}$%
, $\widetilde{g}\bar{E}/E$. Hence the effective critical theory still
contains \emph{two} anisotropic propagators and \emph{six} non-linear
vertices: 
\begin{eqnarray}
&&\partial _{t}\Sigma =\gamma \left( c\,\partial ^{2}+\nabla _{\perp
}^{2}\right) \Sigma +\gamma \,\partial \left( \frac{E}{2}\,\Sigma ^{2}+\frac{%
\bar{E}}{2}\,\Delta ^{2}\right) +\zeta _{\Sigma }\ ,  \label{ads} \\
&&\partial _{t}\Delta =-\lambda \left( r_{\Delta }-a\,\partial ^{2}-\nabla
_{\perp }^{2}\right) \Delta -\lambda \left( \frac{g}{6}\,\Delta ^{3}+\frac{%
\widetilde{g}}{2}\,\Delta \,\Sigma ^{2}\right)  \nonumber \\
&&\qquad +\lambda \left( \widetilde{E}\,\Delta \,\partial \Sigma +\widetilde{%
E}^{\prime }\,\Sigma \,\partial \Delta \right) +\zeta _{\Delta }\ ,
\label{add}
\end{eqnarray}
Similar to the D-S line, all diagrams that contain the conserved noise $%
\zeta _{\Sigma }$ are non-singular. Thus, in contrast to, e.g., model C, $%
\zeta _{\Sigma }$ becomes \emph{irrelevant} here. Instead, the dynamics of
the conserved field is dominated, via the coupling $\bar{E}$, by the
fluctuating, non-conserved, critical $\Delta $. This remarkable feature may
be observable in simulations of the $\Sigma $-$\Sigma $ correlations, which
should develop anomalous (beyond the generic variety \cite{LRC}) long-range
components.

Carrying out an $\epsilon $ ($\equiv 4-d$)-expansion at the one-loop level,
we find 35 non-trivial Feynman graphs, complicated further by the two
distinct anisotropies in $\Delta $ and $\Sigma $. To be brief, we only
report the salient features and leave details to be published elsewhere \cite
{STZ}. The RG fixed points are determined from \emph{seven} coupled RG flow
equations for five marginal non-linearities ($\bar{g}=g/a^{1/2}$, $f=E\bar{E}%
/a^{3/2}$, $\widetilde{f}=\widetilde{E}\bar{E}/a^{3/2}$, $\widetilde{f}%
^{\prime }=\widetilde{E}^{\prime }\bar{E}/a^{3/2}$ and $h=\widetilde{g}\bar{E%
}/Ea^{1/2}$) and two dimensionless ratios ($v\equiv c/a$, $w\equiv \gamma
/\lambda $). All equations involve $\left( v,w\right) $ in algebraic
expressions and, in general, fixed points $\left( v^{*},w^{*}\right) $ are
not simply $O\left( \epsilon \right) $. Aside from the Gaussian fixed point,
the most interesting non-trivial zeroes of the RG $\beta $-functions are the 
\emph{symmetric} fixed points ($\widetilde{f}^{*}=\widetilde{f}^{\prime
*}=(fh/\bar{g}v)^{*}\neq 0$) and the fixed points restoring Galilean
invariance ($h^{*}=0$, $\widetilde{f}^{\prime *}=w^{*}f^{*}$).
Unfortunately, none are stable under the RG flow. The completely \emph{stable%
} fixed points we found are essentially in the class of model A: $\bar{g}%
^{*}=\epsilon /3+O(\epsilon ^{2})$ and $\widetilde{f}^{*}=\widetilde{f}%
^{\prime }{}^{*}=h^{*}=0$. Remarkably, two unusual features emerge: First, $%
(v^{*},w^{*})$ remain undetermined by the flow equations. Instead, they are
constrained by the stability analysis alone, resulting in a stable fixed 
\emph{domain} $\mathcal{D}$ \cite{STZ}. Second, for each $(v^{*},w^{*})\in 
\mathcal{D}$, the coupling $f^{*}$ assumes a non-trivial value $\epsilon
\left[ \frac{3w^{*2}}{(\sqrt{1+v^{*}\,w^{*}}+\sqrt{1+w^{*}})^{2}}-\frac{%
3w^{*}}{(1+\sqrt{v^{*}})^{2}}\right] ^{-1}+O(\epsilon ^{2})$. However, none
of these novel aspects affect the leading exponents, as we summarize below.

Aside from technicalities, the features of the D-FE and D-S lines are
similar: The critical exponents are determined by the critical field alone,
while the noise of the massive mode becomes irrelevant. A single coupling
survives, through which the critical field influences the dynamics of the
massive mode, but not vice versa. Thus, on the D-FE line, we find standard
model A behaviour: $C_{\Delta \Delta }(q,\omega ,r)\,=\,q^{-z-2-\eta }\,{%
\hat{C}}_{\Delta }\left( q\,r_{\Delta }^{-\nu },\,\omega /q^{z}\right) $,
with $\eta =0$, $\nu ^{-1}=2-\epsilon /3$ and $z=2$, up to $O(\epsilon )$
(for $2\leq d\leq 4$) and \emph{trivial} anisotropic properties. Meanwhile,
the leading singularities of the $\Sigma $ field are of the simple diffusive
type: $C_{\Sigma \Sigma }(q,\omega )\,=\,q_{\perp }^{-4}\ {\hat{C}}_{\Sigma
}\left( \omega /q^{2}\right) $ (with no anomalous anisotropy: $\kappa =0$).
We stress, however, that unlike the usual kinetic Ising model, our bi-layer
system exhibits non-vanishing and anomalous \emph{three-point} correlations 
\cite{3pf} --- an obvious signature of the external drive. These should be
clearly detectable in simulations. In addition, corrections to scaling are
expected to be distinct from those of model A.

\emph{Summary and discussion.} To conclude, we find that the universal
critical behaviour of a driven, bilayer Ising lattice gas is controlled by
two distinct universality classes, depending on whether the system orders
into S or FE configurations, i.e., whether the conserved sum, $\Sigma $, or
the non-conserved difference, $\Delta $, of the coarse-grained layer
magnetisations becomes critical. In both cases, the fixed point theory shows
the critical field decoupling completely from the non-critical one, as the
noise of the latter becomes irrelevant. Thus, all along the D-S line, the
critical dimension is $d_{c}^{\Sigma }=5$, and the scaling exponents for the
critical $\Sigma $-field are just those of the ``standard'' DDLG. In
contrast, for the D-FE transitions, the leading singularities are that of
model A for the $\Delta $-field, with $d_{c}^{\Delta }=4$ and isotropic
exponents. At the bicritical point, though both\emph{\ }$\Sigma $ and $%
\Delta $ are critical, the larger critical dimension is $d_{c}^{\Sigma }=5$,
so that the system again displays DDLG-like critical properties. However, at
all critical points, a single non-trivial operator survives which provides a 
\emph{one-way} coupling of the critical field into the dynamics of the
non-critical one. For the D-S transitions, this operator is $\sigma =%
\widetilde{g}c/E\widetilde{E}$, while its counterpart at the D-FE
transitions is $f=E\bar{E}/a^{3/2}$. Interestingly, both have \emph{negative 
}fixed point values, indicating that the three-point couplings do not all
share the same sign. The stability domains of these fixed points are
bounded, so that trajectories starting outside the stable region can run off
to infinity \cite{STZ}. Even for stable trajectories, the presence of
numerous other, unstable fixed points tends to enlarge crossover regimes,
shrouding the asymptotic universal scaling behaviour. This may in fact be at
the core of the inconsistent exponent values observed in Ref.~\cite{jsw}. A
better understanding of the bare couplings, associated with the microscopic
Hamiltonian, would be extremely helpful.

\emph{Acknowledgements.} This research is supported in part by the National
Science Foundation through the Division of Materials Research and the
Jeffress Memorial Trust. RKPZ thanks H.W. Diehl for hospitality at
U-GH Essen, where some of this research was done.

\end{document}